\documentclass[aps,prb,superscriptaddress,reprint,amsmath,amssymb]{revtex4-1}

\usepackage{graphics}
\usepackage[pdftex]{graphicx}
\usepackage{tabularx}
\graphicspath{{./Figures/}}
\usepackage{longtable}
\usepackage{amsmath}
\usepackage{color,soul}
\usepackage{braket}
\usepackage{float} 

\usepackage{multirow}

\begin{document}

\title{Superconducting switching due to triplet component in the Pb/Cu/Ni/Cu/Co$_2$Cr$_{1-x}$Fe$_x$Al$_y$ spin-valve structure}

\author{A.A.~Kamashev}
\affiliation{Zavoisky Physical-Technical Institute, FRC Kazan Scientific Center of RAS, 420029 Kazan, Russia}

\author{N.N.~Garif'yanov}
\affiliation{Zavoisky Physical-Technical Institute, FRC Kazan Scientific Center of RAS, 420029 Kazan, Russia}

\author{A.A.~Validov}
\affiliation{Zavoisky Physical-Technical Institute, FRC Kazan Scientific Center of RAS, 420029 Kazan, Russia}

\author{J.~Schumann}
\affiliation{Leibniz Institute for Solid State and Materials
Research IFW Dresden, D-01069 Dresden, Germany}

\author{V.~Kataev}
\affiliation{Leibniz Institute for Solid State and Materials
Research IFW Dresden, D-01069 Dresden, Germany}

\author{B.~B\"{u}chner}
\affiliation{Leibniz Institute for Solid State and Materials
Research IFW Dresden, D-01069 Dresden, Germany}
\affiliation{Institute for Solid State and Materials Physics, Technical University
Dresden, D-01062 Dresden, Germany}

\author{Ya.V.~Fominov}
\affiliation{L.~D.\ Landau Institute for Theoretical Physics RAS, 142432 Chernogolovka, Russia}
\affiliation{Moscow Institute of Physics and Technology, 141700 Dolgoprudny, Russia}
\affiliation{National Research University Higher School of Economics, 101000 Moscow, Russia}

\author{I.A.~Garifullin}
\affiliation{Zavoisky Physical-Technical Institute, FRC Kazan Scientific Center of RAS, 420029 Kazan, Russia}

\date{\today}

\begin{abstract}
We report the superconducting properties of the
Co$_2$Cr$_{1-x}$Fe$_x$Al$_y$/Cu/Ni/Cu/Pb spin-valve structure which magnetic part comprises the Heusler alloy layer HA = Co$_2$Cr$_{1-x}$Fe$_x$Al$_y$
with a high degree of spin polarization (DSP) of the conduction band and the Ni layer of variable thickness. We obtained that the separation between the superconducting transition curves measured for the parallel ($\alpha = 0^\circ$) and perpendicular
($\alpha = 90^\circ$) orientation of the
magnetizations of the HA and Ni layers reaches up to 0.5\,K ($\alpha$ is an angle between the magnetization of two ferromagnetic layers).
For all studied samples the dependence of the superconducting transition temperature $T_c$ on $\alpha$ demonstrates a deep minimum in the vicinity of the perpendicular configuration of magnetizations. This suggests that the observed minimum and the corresponding full switching effect of the spin valve  is caused by the long-range triplet component of the superconducting condensate in the multilayer. Such a large effect can be attributed to a half-metallic nature of the HA layer which in the orthogonal configuration efficiently draws off the spin-polarized Cooper pairs from the space between the HA and Ni layers. Our results indicate a significant potential of the concept of the superconducting spin-valve multilayer comprising a half-metallic ferromagnet recently proposed by A.~Singh {\it et al.}, Phys.~Rev. X {\bf 5}, 021019 (2015) in achieving large values of the switching effect.  
\end{abstract}



\maketitle

\section{Introduction}

Metallic thin-film heterostructures stay already for decades in the focus
of fundamental research in condensed matter physics and materials science.
They show novel fundamental physical phenomena absent in the initial materials
out of which the heterostructures are made, and, moreover, those phenomena
can also be of remarkable technological relevance. A very prominent example
is the discovery in 1988 of the giant magnetoresistance (GMR) effect
in heterostructures composed of alternating layers of ferromagnetic
and nonmagnetic  metallic layers which opened in fact the new era
of electronics, the so-called spin electronics or, in short, spintronics \cite{Baibich,Grunberg,Prinz}.

A new, more recent development in spintronics was based on the idea
of integrating superconducting layers into the heterostuctures which
gave rise to the field of superconducting (SC) spintronics (for a review see, e.g., Ref.~\cite{Linder}).
The so-called superconducting spin valve (SSV) effect which can be used in SC spintronics was proposed theoretically for the first time by Oh {\it et al.} \cite{Oh} and later on by Tagirov \cite{Tagirov}. The construction suggested by Oh {\it et al.} was the F1/F2/S structure, where F1 and F2 are the ferromagnetic (F) layers and S is the SC layer, whereas Tagirov proposed
a different stacking of the layers F1/S/F2. In both cases the "handle" that switches the SC current in the trilayer on and off is the exchange field from two F layers acting on the S layer. This field is larger for the P orientation of the magnetizations of the F layers (SC is "off") rather than for their AP orientation (SC is "on"). Experimentally, the F1/S/F2 structure was realized first. Gu {\it et al.}~\cite{Gu} found in the system CuNi/Nb/CuNi the magnitude of the SSV effect $\Delta T_c = T_c^{AP}-T_c^P$ (where $T_c^{AP}$ and $T_c^P$ are the SC transition temperatures for the AP and P orientation of magnetizations of the F1 and F2 layers) amounting to 6\,mK and the width of the SC transition curves $\delta T_c \sim 0.1$\,K. Unfortunately, the full switching between the normal and SC states could not be achieved because the necessary relation between $\Delta T_c$ and $\delta T_c$, $\Delta T_c > \delta T_c$ was by far not fulfilled. Since then much experimental work reviewed, e.g., in Refs.\cite{Linder,Blamire,Garifullin}, has been done until in 2010 some of the present authors have demonstrated the full on/off switching between the SC and normal states in the Oh's {\it et al.} type heterostructure Fe1/Cu/Fe2/In with $\Delta T_c=19$\,mK and $\delta T_c \sim 7$\,mK \cite{Leksin2010}. That the F1/F2/S structure is beneficial in achieving the full SSV was previously indicated by the results in Ref.~\cite{Westerholt} were the possible value of $\Delta T_c \sim 200$\,mK in the superlattice [Fe$_2$V$_{11}$]$_{20}$ was obtained indirectly.

One further very remarkable advantage of the F1/F2/S system is its functionality as an SC {\em  triplet} spin valve theoretically predicted by Fominov {\it et al.} \cite{Fominov}. It is related to the generation of the long-range triplet component (LRTC) of the SC condensate at noncollinear orientations of the magnetizations
of the F1 and F2 layers and yields a minimum of the SC critical temperature $T_c$ of the system in an approximately orthogonal geometry. This theoretical prediction was experimentally confirmed for the first time by some of us in the study of the Fe1/Cu/Fe2/Pb multilayer \cite{Leksin2012}. A constantly growing experimental and theoretical interest to the various aspects of the LRTC and its implications for the functionality of SSVs has evolved by now into a new area in the  field of SC spintronics (see, e.g., reviews \cite{Linder,Blamire,Garifullin}).

Recently, Singh {\it et al.} \cite{Aarts2015} have reported a record value of $\Delta
T_c^{trip} \sim 0.6 - 0.8$\,K due to LRTC in the CrO$_2$/Cu/Ni/MoGe heterostructure
where one of the F layers was made of the half-metallic compound CrO$_2$.
Here,  $\Delta T_c^{trip}=T_c
(\alpha=0^\circ)-T_c(\alpha=90^\circ)$ and $\alpha$ is the angle
between the directions of the magnetization of the two F layers.
The reason for the large effect was attributed to the efficiency
of the half-metallic CrO$_2$ layer in drawing off the spin-polarized
Cooper pairs from the space between the two F layers.

The goals of the present work were twofold. First, we considered it necessary to verify the breakthrough results by Singh {\it et al.} \cite{Aarts2015}, and, second, which was even more important, to answer the question whether the proposed concept of the SSV with a half-metallic F element is of a general character, i.e., if a large SSV effect can be realized using materials other than CrO$_2$ in the magnetic part and other than MoGe in the superconducting part of the SSV. Indeed, as will be shown below, we could verify and generalize the results of the pioneering work by Singh {\it et al.} \cite{Aarts2015}. Previously, we have shown the advantages of using the Heusler alloy Co$_2$Cr$_{1-x}$Fe$_x$Al$_y$ as a weak ferromagnet in the F2 layer of the F1/F2/S SSV structure \cite{Kamashev2018}. Therefore, unlike in Ref. \cite{Aarts2015}, instead of CrO$_2$, which in accordance with the data on point contact spectroscopy \cite{Moodera}, has a 90\% polarization of the conduction band, we have chosen as a drawing layer for LRTC the Heusler alloy (HA) Co$_2$Cr$_{1-x}$Fe$_x$Al$_y$ with the spin polarization of the conduction band $\geq 70\%$ \cite{Kamashev2017} and instead of MoGe as an S layer we have used the elemental superconductor Pb.

\section{Sample preparation and experimental results}

We prepared several sets of the F1/F2/S spin-valve structures containing HA = Co$_2$Cr$_{1-x}$Fe$_x$Al$_y$ with the high degree of the spin polarization (DSP) of the conduction band standing for the F1 layer adapting the preparation method from Ref.\cite{Kamashev2017}. The grown heterostructures have the following composition: MgO/Ta(5nm)/HA(20nm)/Cu(4nm)/
Ni($d_{Ni}$)/Cu(1.5nm)/Pb(105nm)/Si$_3$N$_4$ with the variable Ni layer thickness $d_{Ni}$ in the range from 0.6 to 2.5\,nm. In this construction MgO(001)
is a high quality single crystalline substrate, Ta(5nm) is a buffer layer necessary for the optimal growth of the whole structure, HA and Ni play the
roles of the ferromagnetic F1 and F2 layers, respectively, Cu(4nm) decouples the magnetizations of the F1 and F2 layers, Pb(105nm) is an S
layer, Si$_3$N$_4$ is a protective layer against oxidation, Cu(1.5nm) is a buffer layer necessary for the optimal growth of the Pb layer. The Ni, Cu and Pb layers were prepared using e-beam technique. The DC sputtering technique was used for the fabrication of the HA and the Si$_3$N$_4$ layers. We used the deposition rates of 0.4\,{\AA}/s for HA and  Si$_3$N$_4$, 0.5\,{\AA}/s for Cu and Ni, and 10\,{\AA}/s for the Pb films. At first, when evaporating HA, the substrate temperature was kept at $T_{sub}=700$\,K to achieve the desired spin polarization of the HA's conduction band.
Indeed, in accordance with our previous work \cite{Kamashev2017} the composition of our alloy which we  name a Heusler alloy (HA) is, in reality, Co$_2$Cr$_{1-x}$Fe$_x$Al$_{0.63}$ with $x=0.48$. Obviously, there is deficiency of aluminum in this compound in comparison with the ideal Heusler composition Co$_2$Cr$_{1-x}$Fe$_x$Al$_y$. At the same time, in fact, this "not ideal" composition demonstrates a high DSP of the order of 70\% \cite{Kamashev2017}. The study by S. Husain {\it et al.} in Ref. \cite{Husain2016} shows that the DSP increases with increasing the substrate temperature $T_{sub}$. Therefore, we expect the DSP in our samples to be of the order of 80\%.
According to our previous study in Ref.~\cite{Nano}
to improve the smoothness of the Pb layer the substrate temperature should be reduced down to $T_{sub}\sim 150$\,K. Therefore, the top Cu(4nm)/Ni($d_{Ni}$)/Cu(1.5nm)/Pb fragment was grown at this reduced $T_{sub}$. Finally, all samples were covered with a protective Si$_3$N$_4$ layer to prevent oxidation of the Pb layer.

The Ni layer with the  thickness $d_{Ni}\leq 2$\,nm has coercive field of the order of 2 kOe \cite{Leksin2009}. In the present study the Ni layer is deposited at the substrate temperature $T_{sub} \sim 150$ K. Therefore its coercive field should be even larger because the density of dislocation increases with decreasing $T_{sub}$.

As to the HA layer, our SQUID magnetization measurements show that the onset of the saturation of its magnetization occurs at 30\,Oe. At higher magnetic fields the magnetization continues to increase slightly up to the magnetic field of 3\,kOe possibly due to some magnetic inhomogeneity of the HA layer. We note that the magnetic response from the Ni layer cannot be resolved here due to its small value.

The electrical resistivity was measured using the
standard four-point method. The top insulating layer (Si$_3$N$_4$)
was mechanically removed from the areas where the golden wires
should be attached using a silver paste. The quality of the Pb layer
can be judged from the residual resistivity ratio $RRR=R({\rm 300 K})/R({\rm 10 K})=[\rho_{ph}({\rm 300 K})+\rho({\rm 10 K})]/\rho({\rm 10 K})$. Here $R(T)$ is the measured resistance at a given temperature $T$, $\rho({\rm 300K})$ is the phonon contribution to
the specific resistivity at 300\,K, and $\rho({\rm 10 K})$ is the residual resistivity at 10\,K (i.e., above $T_c$). For our samples this ratio amounted
to $RRR= 10 - 12$ which corresponds to the SC coherence length $\xi_s\sim 41 - 45$\,nm (for details see Ref.~\cite{Leksin2015}). The critical temperature $T_c$ is defined as the midpoint of the SC transition curve. Its width in zero magnetic field varied from 20 to 50\,mK dependning on the particular series of the samples and increased with the applied field up to $\sim 250$\,mK (see, e.g., Fig.\,2 below). The narrow SC transition is a characteristic feature of the high quality Pb layer.

\begin{figure}
\begin{center}
\includegraphics[width=0.8\linewidth]{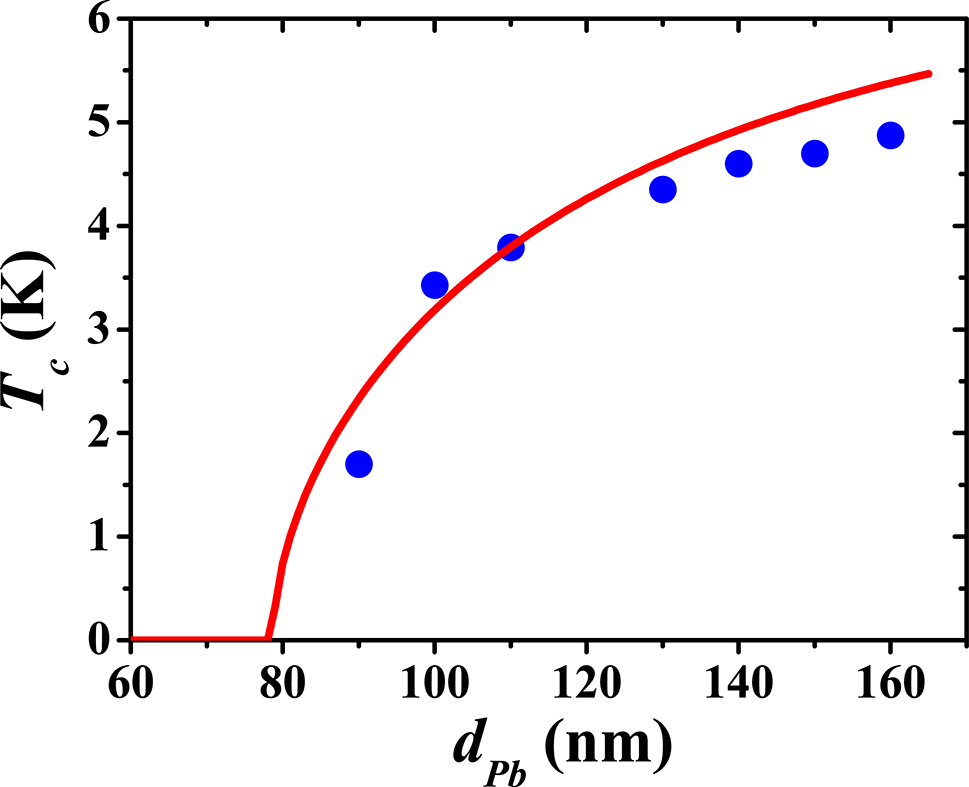}
\end{center}
\caption{(Color online) SC critical temperature $T_c$ vs thickness
of the Pb layer $d_{Pb}$ at a fixed thickness of the Ni layer $d_{Ni}=5$\,nm
for the trilayer Ni(5nm)/Cu(1.5nm)/Pb. Solid line is the theoretical
fit according to Ref.~\cite{Fominov} with the coherence length for the Pb layer  $\xi_S = 41$\,nm. \label{TcPb} }
\end{figure}

For the optimal operation of the SSV it is important to find the optimal thickness of the Pb layer $d_{Pb}$ which should be sufficiently small to make the whole S layer sensitive to the magnetic part of the system. Only in this case the mutual orientation of the magnetizations of the F1 and F2 layers would affect the $T_c$ of the stack. In order to determine the optimal thickness we measured the dependence of $d_{Pb}$ on $T_c$  for the heterostructure
MgO/Ta(5nm)/Ni(5nm)/ Cu(1.5nm)\-/Pb($d_{Pb}$)/Si$_3$N$_4$. Fig.~\ref{TcPb} shows the $T_c(d_{Pb})$ dependence  at a fixed thickness of the Ni layer $d_{Ni}=5$\,nm, which the much larger than the penetration depth of the Cooper pairs in the Ni layer. The $T_c$ decreases rapidly when the thickness of the Pb layer $d_{Pb}$ is reduced down
to 100\,nm. For $d_{Pb}\leq 90$\,nm the value of $T_c \leq 1.5$\,K. Therefore, the optimal thickness range of the Pb layer lies between 90 and 110\,nm and for the further study of the SSV effect we have chosen $d_{Pb} =105$\,nm.

Moreover, this procedure is standard for a simple estimation of the boundary parameters. In particular,  it enables to determine the critical thickness of the SC layer below which superconductivity vanishes $d_s^{crit}$.
From this we obtaine in accordance with Appendix in Ref. \cite{Fominov2002} the transparency parameter of the S/F2 interface $\gamma^b_{SF2} \sim 0.4$.

For the measurements of the angular dependence of $T_c$ in the prepared SSV multilayers we have fixed the magnetization of the F2 layer (Ni) in a certain direction by cooling the sample in magnetic field down to the operational temperatures of the SSV. The magnetization of the F1 layer (HA) can still be easily rotated by an angle $\alpha$ with respect to the pinned magnetization of the Ni layer by external in-plane field. To manipulate the magnetization direction of the HA layer the magnetic field of 30\,Oe should be enough. We performed such experiments and find a disappointingly small SSV effect. Then, just for curiosity, we extended our study to higher magnetic fields. Surprisingly, we found that with increasing the magnetic field the triplet contribution to the SSV effect linearly increases with magnetic field. For example, for the sample PLAK4216 $\Delta T_c^{trip}$ increases linearly up to 0.4 K at 2\,kOe (see, e.g., Fig.\,4 below). 

\begin{figure}
\begin{center}
\includegraphics[width=0.8\linewidth]{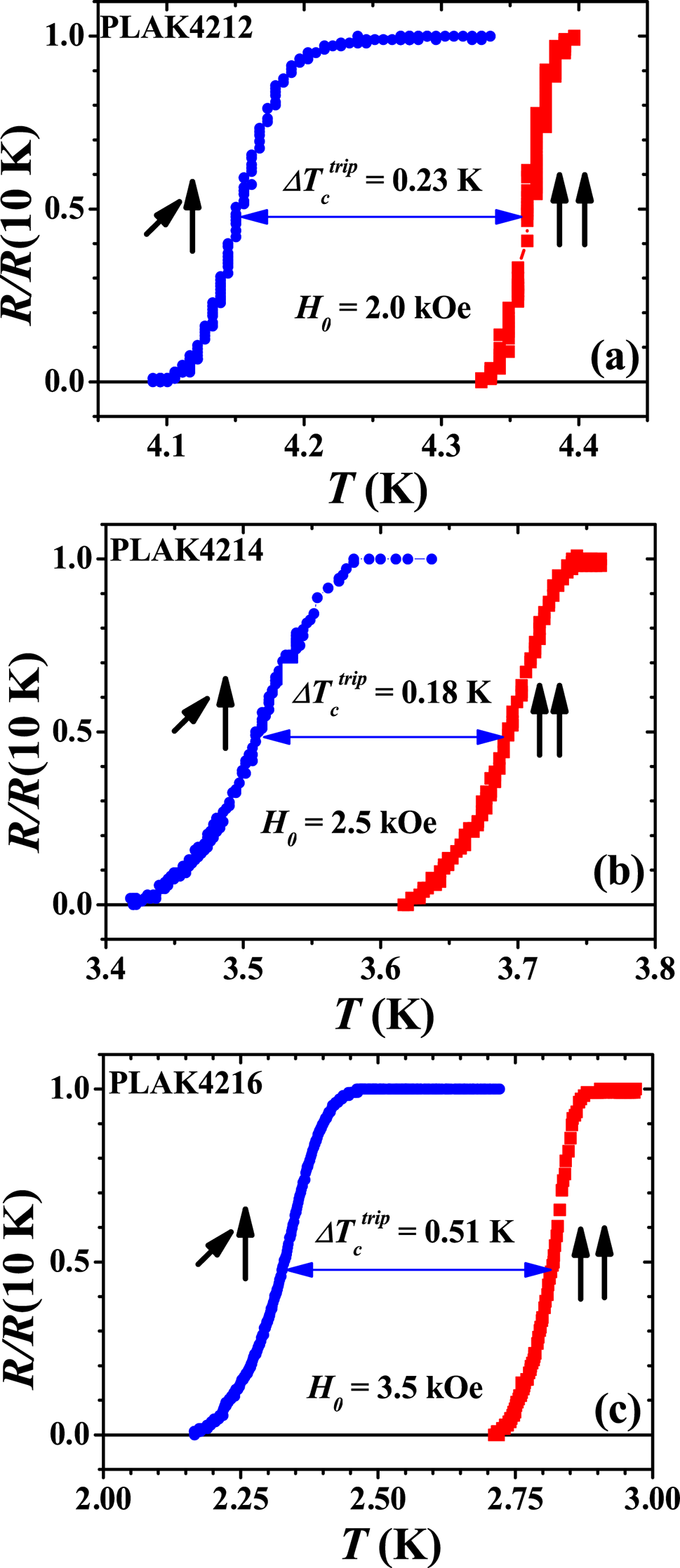}
\end{center}
\caption{SC transition curves for the P and
PP configuration of the cooling field used to fix the direction of the magnetization of the Ni layer and the applied magnetic
field $H_0$ that rotates the magnetization of the HA layer:
(a)~sample PLAK4212 at $H_0=2$\,kOe;
(b)~sample PLAK4214 at $H_0=2.5$\,kOe;
(c)~sample PLAK4216 at $H_0=3.5$\,kOe.}
\label{SCcurves}
\end{figure}

Fig.~\ref{SCcurves} shows the SC transition curves for three representative
samples. The shift of the curves between the P ($\alpha = 0$) and perpendicular (PP) ($\alpha = 90^\circ$) orientations $\Delta T_c^{trip}$ varies between 0.18 and and 0.51 K.

\begin{figure}
\begin{center}
\includegraphics[width=0.8\linewidth]{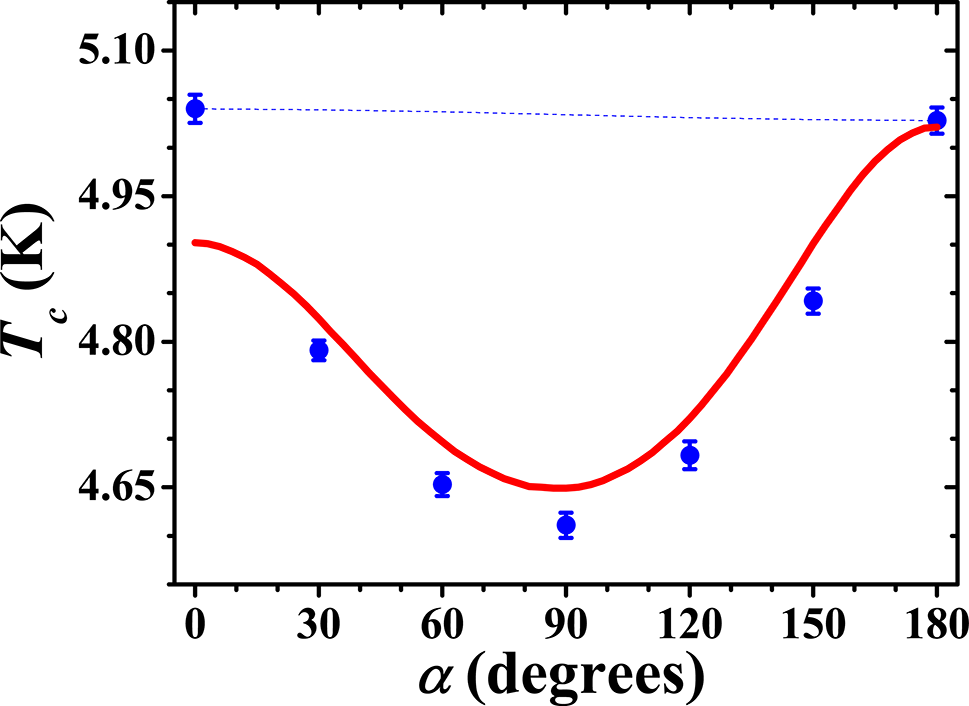}
\end{center}
\caption{Dependence of $T_c$ on the angle $\alpha$ between the
direction of the cooling field  used to fix the direction of the magnetization of the Ni layer and the applied magnetic
field $H_0 =2$\,kOe that rotates the magnetization of the HA layer for
sample PLAK4216 HA(20nm)/Cu(4nm)/Ni(2.5nm)/Cu(1.5nm)/Pb(105nm).
Solid line is the theoretical curve with the parameters presented in
Sect.~Discussion.}\label{TcAngle}
\end{figure}

Fig.~\ref{TcAngle} depicts the  dependence of $T_c$ on $\alpha$ for sample PLAK4216. It appears qualitatively similar to the ones observed by us previously (see, e.g., Refs. \cite{Leksin2012, Garifullin2014a, Garifullin2014b, Garifullin, Leksin2015, Leksin2016a, Leksin2016b, Kamashev2018}), reaching a minimum near $\alpha=90^\circ$. However, the minimum which we observe now is much deeper, suggesting that the SSV effect is dominated by the spin polarized (triplet) Cooper pairs. The main parameters of the studied SSV samples are listed in Table~1.

\begin{table}
\caption{Parameters of all studied samples with the variable Ni-layer
thickness $d_{Ni}$. $\Delta T_c^{trip}$ is the maximum value of the triplet SSV effect as determined from the angular dependence $T_c (\alpha)$  obtained at the field $H_0$. }
\label{tabl1}
\begin{center}
\begin{tabular}{|c|c|c|c|}
\hline Name & $d_{Ni}$, nm & $\Delta T_c^{trip}$, K & $H_0$, kOe \\
\hline PLAK4211 & 0.6 & 0.05 & 1.0 \\ \hline PLAK4212 & 0.9 & 0.23 &
2.0 \\ \hline PLAK4213 & 1.3 & 0.13 & 2.0 \\ \hline PLAK4214 & 1.6 &
0.18 & 2.5 \\ \hline PLAK4215 & 2 & 0.05 & 1.25 \\ \hline PLAK4216 &
2.5 & 0.51 & 3.5 \\ \hline
\end{tabular}
\end{center}
\end{table}

The magnitude of the triplet SSV effect $\Delta T_c^{trip}$ depends practically linearly on $H$ at small applied magnetic fields up to the field $H_0$ which values are listed in Table~\ref{tabl1} together with the corresponding values of $\Delta T_c^{trip}$.

\begin{figure}
\begin{center}
\includegraphics[width=0.8\linewidth]{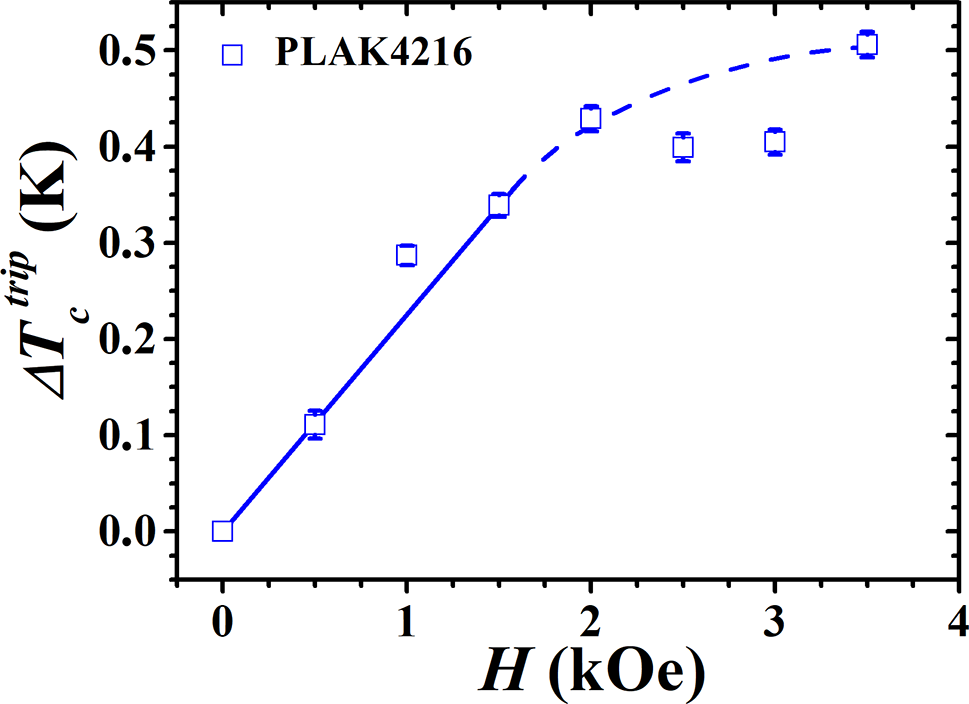}
\end{center}
\caption{(Color online) The magnitude of the triplet SSV vs applied magnetic field for sample PLAK4216. Line is guide for eyes.}\label{TripEff}
\end{figure}

At first glance it is surprising that
$\Delta T_c^{trip}$ increases well above the
saturation magnetic field for the HA layer. We suppose that this may be caused by some magnetic inhomogeneity of the HA layer reflected in a slight increase of its magnetization up to the field of 3\,kOe, where more and more "microdomains" become gradually involved in the formation of the total moment. Fig.\ref{TripEff} shows the dependence of the triplet contribution to the SSV effect on the external magnetic field for the sample PLAK4216. Notably, a similar increase of $\Delta T_c^{trip}$ was observed by Singh {\it et al.} \citep{Aarts2015} as well, for which no conclusive explanation can be found at present. Obviously, this field dependent effect observed by two groups on different samples appears to be a salient feature of the new type of SSVs and needs theoretical explanation.

\section{Discussion}\label{discus}

A remarkably large separation of the SC transition curves for the P and PP
orientation of magnetizations of F1 and F2 layers yielding the value of  $\Delta T_c^{trip}$ up to 0.5\,K in not too strong magnetic fields as compared to those used in Ref.~\cite{Aarts2015} evidences prominent spin-triplet superconducting
correlations in our samples. The theoretical approach that we employ for the analysis of experiment in Fig.~\ref{TcAngle}, is based on generalization of Ref.~\cite{Fominov} along the lines of Refs.~\cite{Deminov2015,Deminov2015a}. It allows to consider layered structures with different material parameters of all the layers and arbitrary Kupriyanov-Lukichev boundary parameters \cite{Kupriyanov1988} of all the interfaces.

Each of the two interfaces (F/S and F/F) is described by the matching parameter $\gamma$ and the resistance parameter $\gamma_b$ \cite{Deminov2015}:
\begin{gather}
\gamma_{FS} = \frac{\rho_S \xi_S}{\rho_{F2} \xi_{F2}},
\qquad
\gamma_{bFS} = \frac{R_{bFS} \mathcal A}{\rho_{F2} \xi_{F2}},
\\
\gamma_{FF} = \frac{\rho_{F2} \xi_{F2}}{\rho_{F1} \xi_{F1}},
\qquad
\gamma_{bFF} = \frac{R_{bFF} \mathcal A}{\rho_{F1} \xi_{F1}}.
\end{gather}
Here, $\rho$ and $\xi$ are the resistance and the coherence length of the layers, $R_b$ and $\mathcal A$ are the interface resistance and area. The necessity to consider arbitrary F/F interface parameters is due to different materials of the two F layers. This is a new theoretical ingredient, in comparison to fittings of our previous experiments in Refs.~\cite{Leksin2012,Leksin2015,Garifullin}.

Fig.~\ref{TcAngle} demonstrates that theory correctly reproduces characteristic features of the $T_c(\alpha)$ dependence (triplet spin-valve behavior).

Parameters used for fitting of the theory to the experimental
results are the following: coherence length in S-layer $\xi_S = 41$\,nm,
coherence lengths in F2- and F1-layers $\xi_{F2} = 6.25$\,nm and $\xi_{F1} = 40$\,nm. The boundary conditions S/F and F/F
interfaces where the material matching parameter $\gamma$ and the transparency parameter
$\gamma^b$ are $\gamma_{SF} = 0.1$, $\gamma^b_{SF} = 0.1$, $\gamma_{FF} = 1$ and $\gamma^b_{FF} = 0.1$.
The exchange energies of the F1 and F2 layers amount to $h_2 = 0.03$\ eV and $h_1 = 0.39$\ eV.

\section{Conclusions}

By studying the SSV  multulayers Co$_2$Cr$_{1-x}$Fe$_x$Al$_y$/Cu/Ni/Cu/Pb which magnetic part contains the Heusler alloy Co$_2$Cr$_{1-x}$Fe$_x$Al$_y$ with a high degree of spin polarization of the condiction band
we have obtained a large SSV effect due to the long-range triplet component of the superconducting condensate   $\Delta T_c^{trip} \sim  0.5$\,K at a moderate applied field of 3.5\,kOe as compared with the earlier work in Ref.~\cite{Aarts2015}.  Our results show that
there is a potential to achieve large values of $\Delta T_c^{trip}$ even at smaller fields relevant for applications by careful design and optimization of all elements of the SSV heterostructure regarding both the superconducting and magnetic parts. Our observations suggest that that the concept of the SSV with a half-metallic ferromagnetic element proposed in Ref.~\cite{Aarts2015} is of general character. In particular, finding for this purpose the most appropriate ferromagnet with a high degree of spin polarization of the conduction band appears to be a crucial issue. Furthermore, noting first theoretical attempts in Refs.~\cite{Mironov,Halterman},  our data as well as the results by Singh {\it et al.} \cite{Aarts2015} call for a comprehensive quantitative theoretical treatment to obtain further insights into exciting physics of the triplet superconducting spin valves.


%

\end{document}